\documentclass[conference]{IEEEtran}
\IEEEoverridecommandlockouts

\usepackage{float}
\usepackage{cite}
\usepackage{amsmath,amssymb,amsfonts}
\usepackage{algorithm}
\usepackage{algpseudocode}
\usepackage{graphicx}
\usepackage{textcomp}
\usepackage{xcolor}
\usepackage{graphics}
\usepackage{mathrsfs}
\usepackage{epstopdf}
\usepackage{caption}
\def\BibTeX{{\rm B\kern-.05em{\sc i\kern-.025em b}\kern-.08em
    T\kern-.1667em\lower.7ex\hbox{E}\kern-.125emX}}
\allowdisplaybreaks[3]
\makeatletter
\renewcommand{\maketag@@@}[1]{\hbox{\m@th\normalsize\normalfont#1}}%
\makeatother

\begin{document}

\title{Distributionally Robust Physical-Layer Security\\for Satellite Communication via Aerial\\Reconfigurable Intelligent Surface}

\author{
Zhaole Wang,
Xiao Tang,
Naijin Liu,
Jinxin Liu,
Qinghe Du,
Lei Chen,
and Tingwu Lin
\thanks{Z. Wang is with the School of Future Technology, Xi'an Jiaotong University, Xi'an 710049, China.}
\thanks{X. Tang is with the School of Information and Communication Engineering, Xi'an Jiaotong University, Xi'an 710049, China, and also with Shenzhen Research Institute of Northwestern Polytechnical University, Shenzhen 518057, China.}
\thanks{N. Liu is with the Institute of Telecommunication and Navigation Satellites, China Academy of Space Technology, Beijing 100094, China.}
\thanks{J. Liu is with the School of Mechanical Engineering, Xi’an Jiaotong University, Xi’an 710049, China.}
\thanks{Q. Du is with the School of Information and Communication Engineering, Xi'an Jiaotong University, Xi'an 710049, China.}
\thanks{L. Chen and T. Lin are with ZTE Corporation, Shenzhen 518063, China.}
}

\maketitle

\begin{abstract}
Satellite communications are envisioned as a key enabler for ubiquitous coverage in future 6G networks, yet the broadcast nature renders them vulnerable to eavesdropping, especially given the long-distance transmissions and associated high uncertainties. In this paper, we propose the physical layer security enhancement for multi-beam satellite communications with the assistance of an aerial reconfigurable intelligent surface (ARIS). Considering the high dynamics and uncertainties of channels, we characterize the channel distribution with moment-based ambiguity sets. Accordingly, a distributionally robust secrecy rate optimization is formulated through joint design of transmit and reflection beamforming. We then introduce a conditional value-at-risk-based reformulation to convert the probabilistic constraints into deterministic forms. An alternating optimization framework is subsequently employed to iteratively update the transmit and reflective beamforming vectors until convergence. Simulation results demonstrate that the proposed distributionally robust scheme significantly enhances secrecy performance, and maintains reliable performance across various channel error distributions.
\end{abstract}

\begin{IEEEkeywords}
Reconfigurable intelligent surface, multibeam satellite systems, physical layer security, distributionally robust optimization.
\end{IEEEkeywords}

\section{Introduction}
With the integration of Non-Terrestrial Networks (NTN) into 5G-Advanced and future 6G standards, satellite communication is experiencing rapid development, driven by its capability for global coverage and rapid deployment in emergency scenarios~\cite{1,2}. However, the broadcast nature of wireless channels exposes the satellite downlinks to severe security challenges, while traditional encryption schemes face significant challenges related to dynamic key management and system resource consumption~\cite{3,4}. In contrast, Physical Layer Security (PLS) leverages the intrinsic characteristics of wireless channels to achieve secure communication without the need for complex key management~\cite{5,20}. Characterized by low computational overhead and compatibility with existing communication architectures, the PLS is regarded as a promising solution for safeguarding future satellite communication systems~\cite{6}.

Despite the potential advantages of PLS, the severe path loss and high channel correlation in the satellite-ground link constrain its performance in practical satellite communications~\cite{7}. Recently, the emergence of reconfigurable intelligent surfaces (RIS) has provided a new solution for improving PLS in satellite communications~\cite{9}. RIS is composed of a large number of low-cost passive reflective elements, which can be adjusted by controlling the phase shift of each element to achieve regulation of the wireless environment~\cite{10}. Based on the above characteristics, existing studies have explored RIS-assisted satellite secure transmission from different aspects. For the problem of channel correlation, in~\cite{13}, the authors artificially introduce channel differences using RIS and effectively guarantee the performance of PLS. In~\cite{14}, the authors introduce active RIS to address the long-distance transmission loss problem of satellite communication, significantly enhancing the security performance compared with the traditional passive RIS. In~\cite{15}, the authors extend the research scenario to the integrated satellite-terrestrial network and propose a joint optimization strategy of RIS passive beamforming and artificial noise, which maximizes the satellite secrecy rate while meeting the quality of service. Furthermore, considering that it is difficult to obtain perfect channel state information (CSI) in practical scenarios, in~\cite{a1}, the authors investigate the situation of imperfect CSI and propose a hybrid secure transmission scheme for RIS-assisted multi-beam satellite communication.

Although optimizing RIS passive beamforming can improve security performance, its effectiveness is severely constrained by the deployment location of RIS~\cite{16}. Specifically, the terrestrial RIS is difficult to adapt to the high-speed movement of satellites and the random distribution of eavesdroppers, which seriously restricts the PLS performance of the system. To address these challenges, some researchers have taken advantage of the lightweight characteristics of RIS and installed it on aerial platforms such as drones or balloons~\cite{18,19}. By optimizing the deployment location of drones, a reliable line-of-sight (LoS) link can be established, thereby improving the security performance of the system~\cite{c1}. In~\cite{21}, the authors explore a strategy to mitigate malicious attacks by jointly optimizing passive beamforming and flight trajectories in response to the anti-interference issue in air-ground integrated scenarios. Motivated by the significant security gains of ARIS in ground networks, researchers have further explored its potential in ensuring the security of satellite communications. In~\cite{22}, the authors mounted RIS on the unmanned aerial vehicle (UAV) platform and significantly enhanced the PLS performance of the cognitive NTN by jointly optimizing the UAV trajectory and the RIS reflection coefficient. In addition, in~\cite{23}, the authors explore the multi-ARIS collaborative scenario, aiming to enhance the PLS performance of satellite communications by optimizing the deployment location and the association mechanism between multi-ARIS and ground user groups.

It is worth noting that the above mentioned works are mainly based on the assumption that the channel state information (CSI) can be perfectly obtained. However, acquiring accurate CSI remains a significant challenge, particularly regarding the eavesdropper's links~\cite{24}. Existing robust RIS-assisted secrecy studies can generally be classified into two categories. One approach employs the worst-case design based on a bounded-error model, assuming that the CSI error of the eavesdropper is within a bounded sphere~\cite{25,26}. However, this approach often results in overly conservative transmission schemes, and obtaining precise error bounds is challenging in practice. Another approach focuses on outage-based probabilistic methods which rely on specific statistical assumptions regarding CSI errors. For instance, these methods model the error with distributions such as the circularly symmetric complex Gaussian distribution~\cite{27}, aiming to provide QoS outage probability guarantees. However, in practical satellite communication scenarios, the CSI error distribution is often unknown and easily deviates from the pre-assumed model due to high satellite mobility and ARIS jitter. Once the preset distribution is inconsistent with the actual situation, the security performance of the system will face degradation.

However, Distributionally Robust Optimization (DRO) provides a new solution to address the mismatch between the actual CSI and the pre-assumed model. Different from traditional CSI error models, DRO does not rely on prior knowledge of the exact error distribution. Instead, it constructs an ambiguity set based on the statistical information of CSI errors and optimizes system performance under the worst-case distribution within this set. Recently, DRO has attracted increasing attention in terrestrial RIS-assisted networks for handling channel uncertainties. In~\cite{29}, the authors proposed DRO to mitigate localization errors in RIS-assisted distributed MIMO systems, effectively improving the user sum-rate under location uncertainty. Meanwhile, in~\cite{30}, the authors proposed DRO to enhance the physical layer security of terrestrial RIS-aided multicast systems, proving its superiority over non-robust and worst-case designs. However,~\cite{30} primarily focuses on scenarios where a single common data stream is broadcast, avoiding the challenges of inter-beam interference and the presence of multiple users and eavesdroppers. Consequently, despite these successes in terrestrial networks, the potential of DRO in satellite communications has not been fully explored.

In view of the advantages of DRO in improving the robustness of communication systems against channel uncertainty, this paper aims to leverage DRO to enhance the secrecy performance of ARIS-assisted multi-beam satellite communications. The main contributions of this paper are summarized as follows:
\begin{itemize} 
\item We consider a distributionally robust ARIS-assisted secure satellite communication framework that enhances physical layer security by jointly exploiting the transmission and reflection beamforming. We formulate a joint beamforming optimization problem to maximize the sum rate of legitimate users while satisfying distributionally robust outage probability constraints on wiretap rates. Rather than assuming specific error distributions, the security constraints rely only on the statistics of CSI errors, which ensures robustness against worst-case scenarios. 
\item We transform the intractable semi-infinite chance constraints into tractable forms. Specifically, we use Conditional Value-at-Risk (CVaR) to reformulate the probabilistic constraints into convex Linear Matrix Inequalities (LMIs), which ensures that the secrecy outage probability remains below a pre-assumed threshold for any error distributions within the ambiguity set.
\item To address the intractability of the formulated non-convex problem with semi-infinite chance constraints, we develop an alternating optimization (AO) framework to jointly optimize the transmission beamforming and the passive beamforming. Specifically, semidefinite relaxation (SDR) is applied to optimize the transmission beamforming, while a scheme that combines fractional programming (FP) and the penalty-based convex–concave procedure (CCP) is adopted to handle the passive beamforming subject to unit-modulus constraints.
\item Simulation results under various channel error distributions are presented to demonstrate the effectiveness of the proposed method. The results show that the proposed DRO-based design is more robust and achieves better performance in secure transmission than non-robust baselines and schemes without ARIS.
\end{itemize}

The remainder of this article is organized as follows. Section \ref{s2} presents the system model for the ARIS-assisted satellite network. Section \ref{s3.1} formulates the joint optimization problem and presents the decomposition strategy. Section \ref{s3.2} details the optimization algorithms for transmission and passive beamforming. Section \ref{s4} provides the simulation results to verify the robustness and effectiveness of the proposed scheme. Finally, Section \ref{s5} concludes this paper.

\textit{Notations}: Vectors and matrices are represented by bold letters. The transpose and conjugate transpose of a matrix $\boldsymbol{A}$ are represented by $\boldsymbol{A}^{\mathrm{T}}$ and $\boldsymbol{A}^{\dagger}$. For a square matrix $\boldsymbol{A}$, $\mathrm{Tr}(\boldsymbol{A})$ is the trace of $\boldsymbol{A}$. The expression $\mathrm{diag}(a)$ is the diagonal matrix with the elements of $a$, and also $\mathrm{diag}(\boldsymbol{A})$ returns a diagonal matrix with the elements of $\boldsymbol{A}$ on its main diagonal. For a complex scalar number $a$, $|a|$ is used for the absolute value of $a$ and $\Re(a)$ denotes the real part of a complex scalar $a$. For a matrix $\boldsymbol{A}$, $||\boldsymbol{A}||$ denotes the spectral norm of $\boldsymbol{A}$, $\mathrm{rank}(\boldsymbol{A})$ implies the rank of $\boldsymbol{A}$ and $\boldsymbol{A}\succeq0$ represents a positive semi-definite matrix $\boldsymbol{A}$. The expression $\mathcal{O}(\cdot)$ implies the big-O notation.

\section{System Model}\label{s2}
\begin{figure}[t!]
\centering
    \includegraphics[width=0.9\linewidth]{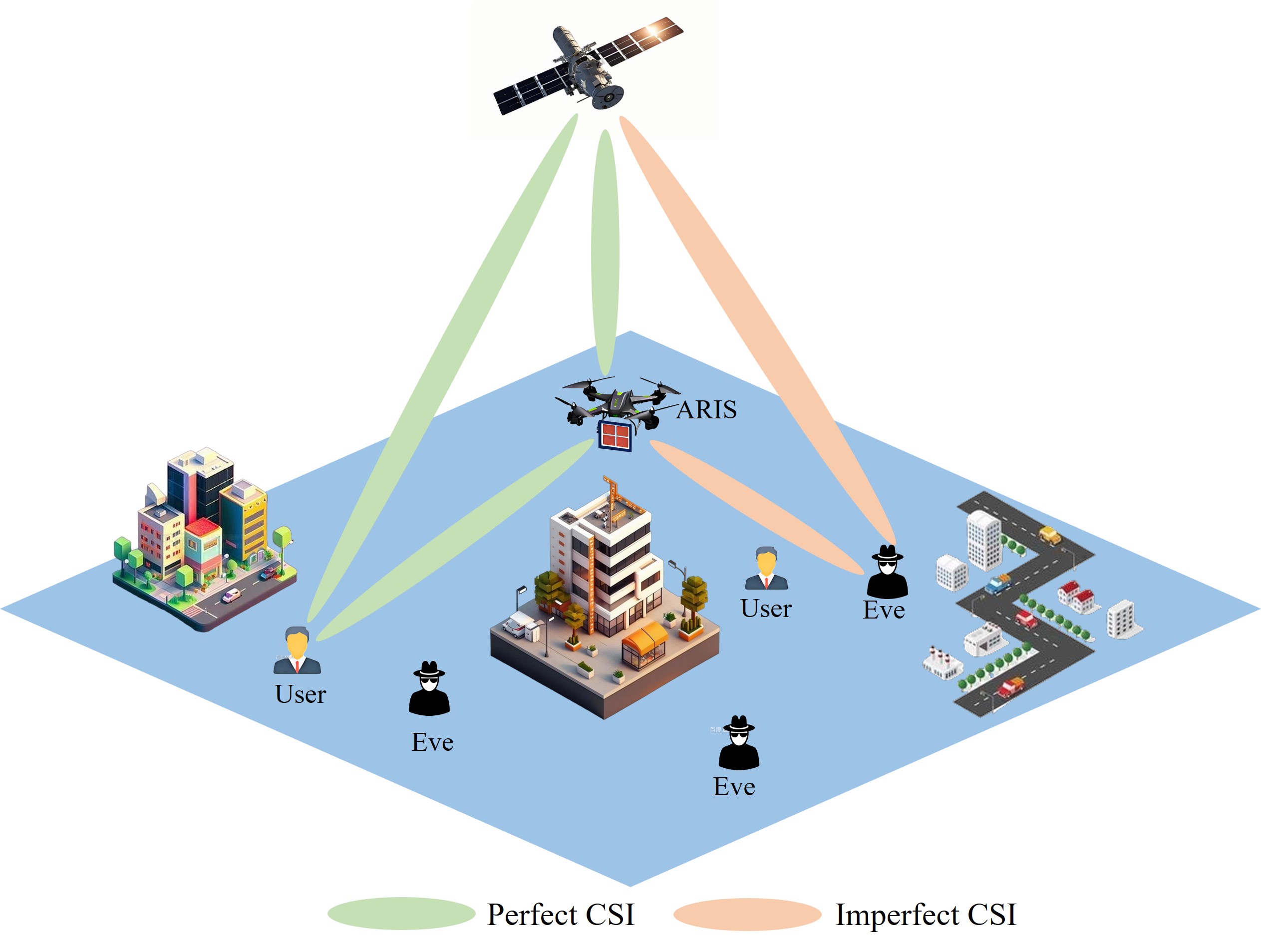}
    \caption{ARIS assisted satellite communication}
    \label{system}
\end{figure}
As shown in Fig. \ref{system}, we consider a multibeam satellite system consisting of $K$ legitimate users. Additionally, there are $E$ eavesdroppers in the system attempting to intercept the legitimate signals. Therefore, the transmission of legitimate users requires secure mechanisms to prevent eavesdropping.  The sets of intended and unintended users are denoted as $\mathcal{K}=\{1,\ldots,K\}$ and $\mathcal{E}=\{1,\ldots,E\}$, where the collection of all ground users can be denoted by $\mathcal{I}=\mathcal{K}\cup\mathcal{E}$. Meanwhile, the multibeam satellite is equipped with $L$ antennas, and the ground user $i\in\mathcal{I}$ is equipped with a single antenna. The two-dimensional horizontal coordinates of the ground user $i\in\mathcal{I}$ are denoted by $\boldsymbol{\omega}_{i}=[x_{i},y_{i}]$, and it is located within an area defined by $\mathcal{X}\times\mathcal{Y}$ with $\mathcal{X}$ and $\mathcal{Y}$ denoting the range along the x-axis and y-axis, respectively.

For the considered scenarios, to address the limitations of terrestrial RIS, we propose to deploy aerial platforms equipped with RIS. In this paper, we assume the ARIS is mounted on a hovering aerial platform, which maintains a fixed position during the transmission. The distance between the ARIS and the ground user $i$ is denoted by $d_{a,i}$. To enhance the reflection-based transmissions, we introduce the concept of subsurface that each ARIS comprises a number of adjacent elements inducing an identical phase shift to the incident signal~\cite{12}. The ARIS consists of $M$ subsurfaces, and each subsurface contains multiple reflective elements with the same phase shift. We denote the phase shifts of the subsurfaces as $\boldsymbol{\theta}=[\theta_{1},\theta_{2},\ldots,\theta_{M}]^{\mathrm{T}}$, then the reflection-coefficient matrix can be given as $\boldsymbol{{\Theta}}=\mathrm{diag}(\boldsymbol{\vartheta})$ with $\boldsymbol{\vartheta}=[e^{i\theta_{1}},e^{i\theta_{2}},\ldots,e^{i\theta_{M}}]$. In addition, we assume the ARIS is cooperatively controlled via onboard processing with reliable backhaul links.

\subsection{Channel Model}
According to the signal propagation characteristics of the LEO satellite communications~\cite{36}, the channel from the satellite to the ground user $i$ is expressed as $\boldsymbol{h}_{i}=\bar{h}_{i}\cdot\boldsymbol{r}^{-\frac12}\odot\boldsymbol{b}^{\frac12}$, and $\boldsymbol{h}_{i}\in\mathbb{C}^{L\times1}$. The free space loss and phase of the channel are defined by  $\bar{h}_{i}=\frac\lambda{4\pi d_{s,i}}e^{-i\frac{2\pi}\lambda d_{s,i}}$, where ${\lambda}$ is the carrier wavelength, and $d_{s,i}$ is the distance between the satellite and the ground user. The rain attenuation $\boldsymbol{r} = \xi ^2 \boldsymbol{1} _{L\times 1}$ follows a log-normal distribution with $\mathrm{ln}(\xi_\mathrm{dB})\sim$ $\mathcal{CN}(\mu,\sigma^{2})$, where $\mu$ and $\sigma$ are the mean value and variance which are related to the satellite communication frequency, polarization mode and the location of the user, respectively. Besides, the beam gain $\boldsymbol{b} = [ b_1,\ldots, b_L]^\mathrm{T}$ mainly depends on the satellite transmit antenna radiation pattern and the user location, where the gain is given by $b_l=G_{l,max}\left(\frac{J_1(u_l)}{2u_l}+36\frac{J_3(u_l)}{u_l^3}\right)^2$, with $u_l=2.07123\sin\varphi_{l}/\sin\left(\varphi_{\mathrm{3dB}}\right)_{l}$, where $G_{l,max}$ is the maximum beam gain for the $l$-th beam, $\varphi_\mathrm{3dB}$ is the angle corresponding to the 3 dB power loss from the beam center.  Additionally, $J_1$ and $J_3$ are the first-kind Bessel functions with first order and third order, respectively.

The channel from the satellite to the ARIS is denoted by $\boldsymbol{G}=[\boldsymbol{G}_{1};
\dots;\boldsymbol{G}_{M}]$, and $\boldsymbol{G}\in\mathbb{C}^{M\times L}$. The channel gain between the satellite and the $m$-th subsurface of the ARIS is given by $\boldsymbol{G}_m=\bar{G}_m\cdot\boldsymbol{r}^{-\frac12}\odot\boldsymbol{b}^{\frac12}$. Consider the ARIS forming a uniform linear array along x-axis, the free space loss and phase can be expressed as $\bar{G}_m=\frac\lambda{4\pi d_{s,a}}e^{-j\frac{2\pi}\lambda d_{s,a}}e^{-j\frac{2\pi d}\lambda(m-1)\phi}$, where $d_{s,a}$ is the distance between the satellite and the ARIS, $d$ is the separation between two adjacent reflecting elements, $\phi$ denotes the cosine of angle of arrival. The rain attenuation is expressed by  $\boldsymbol{r} \in \mathbb{R} ^{1\times L}$, and the beam gain is denoted by $\boldsymbol{b} \in \mathbb{R} ^{1\times L }$.

The ARIS-to-ground transmission experiences both path loss and small-scale fading. The path loss from ARIS to the ground user is given by $C_L(d_{a,i})=L_0d_{a,i}^{-\beta}$, where $L_0$ denotes the path loss at the reference distance of 1 m, $d_{a,i}$ denotes the distance between the ARIS and the ground user $i$, $\beta$ is the path loss exponent for the wireless transmissions on the ground. As for small-scale fading, the Rician fading model is adopted. Therefore, the ARIS-to-ground channel is represented as $\boldsymbol{g}_{i}=\sqrt{C_L(d_{a,i})}(\sqrt{\frac\rho{\rho+1}}\boldsymbol{g}_{i}^\mathrm{LoS}+\sqrt{\frac1{\rho+1}}\boldsymbol{g}_{i}^\mathrm{NLoS})$,
where $\rho$ is the Rician factor, $\boldsymbol{g}_{i}^\mathrm{LoS}=e^{-i\frac{2\pi}{\lambda}d_{a,i}}[e^{-i\frac{2\pi d}{\lambda}(m-1)\phi_{i}}]_{m\in\mathcal{M}}$ denotes the deterministic LoS component which is related to the angle of departure $\phi_{i}=\frac{x_i-x}{d_{i}}$ and $\boldsymbol{g}_{i}^\mathrm{NLoS}$ denotes the non-LoS component modeled as Rayleigh fading with $\boldsymbol{g}_{i}^\mathrm{NLoS}\sim\mathcal{CN}(0,1)$.

\subsection{Signal Model}
For the considered system model, the received direct and reflected signals at the ground user $i$ are given as
\begin{equation}
\begin{aligned}\label{eq1}
y_{i}=\left(\boldsymbol{h}^{\dagger}_{i}+\boldsymbol{g}^{\dagger}_{i}\boldsymbol{\Theta}\boldsymbol{G}\right)\sum_{l=1}^K\boldsymbol{w}_{l}s_{l}+n_{i},\quad\forall i\in\mathcal{I},
\end{aligned}    
\end{equation}
where $\boldsymbol{w}_{l}\in\mathbb{C}^{L\times 1}$ denotes the beamforming vector for the $l$-th legitimate user, along with the transmitted symbol $s_{l}\sim\mathcal{CN}(0,1)$, and $n_{i}$ is the white Gaussian noise at the ground user $i\in\mathcal{I}$. Additionally, the satellite beamforming is constrained by the maximum satellite power $P_T$, which is expressed by $\sum\limits_{k=1}^K\|\boldsymbol{w}_k\|^2\leq P_T$. Then, we obtain the SINR at the $k$-th legitimate user and at the $e$-th eavesdropper with respect to the $k$-th legitimate user as
\begin{align}
\gamma_{k}=\frac{\left|\left(\boldsymbol{h}^{\dagger}_{k}+\boldsymbol{g}^{\dagger}_{k}\boldsymbol{\Theta}\boldsymbol{G}\right){\boldsymbol{w}}_{k}\right|^{2}}{\sum\limits_{l=1,l\neq k}^K\left|\left(\boldsymbol{h}^{\dagger}_{k}+\boldsymbol{g}^{\dagger}_{k}\boldsymbol{\Theta}\boldsymbol{G}\right){\boldsymbol{w}}_{l}\right|^{2}+\sigma_{k}^{2}},\quad k\in\mathcal{K},
\end{align}
\begin{align}
\gamma_{e,k}=\frac{\left|\left(\boldsymbol{h}^{\dagger}_{e}+\boldsymbol{g}^{\dagger}_{e}\boldsymbol{\Theta}\boldsymbol{G}\right){\boldsymbol{w}}_{k}\right|^{2}}{\sum\limits_{l=1,l\neq k}^K\left|\left(\boldsymbol{h}^{\dagger}_{e}+\boldsymbol{g}^{\dagger}_{e}\boldsymbol{\Theta}\boldsymbol{G}\right){\boldsymbol{w}}_{l}\right|^{2}+\sigma_{e}^{2}},\quad k\in\mathcal{K},e\in\mathcal{E},
\end{align}
respectively, where the noise power for a ground user $i$ is expressed as $\sigma_{i}^2=k_B B T_{i}$, where $k_B=1.380649\times10^{-23}J/K$ is the Boltzmann constant, $B$ is the operating bandwidth, and $T_{i}$ is the noise temperature. Then, the achievable rates of all ground users, including the rate of the $k$-th legitimate user and the eavesdropping rate of the $e$-th eavesdropper for the $k$-th legitimate user can be written as
\begin{equation}\label{eq29}
    R_{k}=\log(1+\gamma_{k}),\quad\forall k\in\mathcal{K},
\end{equation}
and
\begin{equation}
R_{e,k} =\log(1+\gamma_{e,k}),\quad\forall e\in\mathcal{E},\forall k\in\mathcal{K},
\end{equation}
respectively.

\subsection{CSI Uncertainty Model}
In the system considered, there are significant differences in the ability to acquire CSI for different links. The ARIS and legitimate users can cooperate with the satellite to collect CSI. Therefore, we assume that the CSI of the satellite-ARIS-legitimate user and satellite-legitimate user links, i.e., $\boldsymbol{G}$, $\{\boldsymbol{g}_k\}_{k=1}^K$, and $\{\boldsymbol{h}_k\}_{k=1}^K$, are perfectly known. For the eavesdropper link, due to its non-cooperative nature, it is difficult for the satellite to obtain its precise CSI. Hence, to characterize this inherent uncertainty, we adopt an imperfect CSI model to describe the eavesdropping channel and consider the following random CSI error model:
\begin{align}
    \boldsymbol{h}_e=\bar{\boldsymbol{h}}_e+\Delta \boldsymbol{h}_e,\quad e\in\mathcal{E},\label{1.1}\\
\boldsymbol{g}_e=\bar{\boldsymbol{g}}_e+\Delta \boldsymbol{g}_e, \quad e\in\mathcal{E},\label{1.1.1}
\end{align}
where $\bar{\boldsymbol{h}}_e$ and $\bar{\boldsymbol{g}}_e$ are the estimated values of $\boldsymbol{h}_e$ and $\boldsymbol{g}_e$ at the satellite, respectively. In addition, $\Delta \boldsymbol{h}_e$ and $\Delta \boldsymbol{g}_e$ denote the corresponding CSI errors. These errors follow distributions over the measurable spaces $\mathbb{C}^L$ and $\mathbb{C}^M$, with mean vectors of $\boldsymbol{\xi}_{h_e}\in\mathbb{C}^{L\times 1}$ and $\boldsymbol{\xi}_{g_e}\in\mathbb{C}^{M\times 1}$, and covariance matrices of $\boldsymbol{\Sigma}_{h_e}\in\mathbb{H}_{+}^{L}$ and $\boldsymbol{\Sigma}_{g_e}\in\mathbb{H}_{+}^{M}$, respectively. In practical satellite systems, the precise error distribution of CSI is often difficult to obtain, but its mean and covariance can still be reliably estimated. Therefore, we do not assume a specific distribution for the error, but adopts a moment-based ambiguity set to characterize the error. Specifically, the CSI error terms $\Delta\boldsymbol{h}_e$ and $\Delta\boldsymbol{g}_e$ are considered to belong to a set of distributions that includes all distributions with a specific mean and covariance, i.e.,
\begin{align}
    \Delta\boldsymbol{h}_e\sim\Delta F_{h_e}\in\mathscr{D}(\boldsymbol{\xi}_{h_e},\boldsymbol{\Sigma}_{h_e}),\label{1.2}\\
      \Delta\boldsymbol{g}_e\sim\Delta F_{g_e}\in\mathscr{D}(\boldsymbol{\xi}_{g_e},\boldsymbol{\Sigma}_{g_e}).\label{1.2.1}
\end{align}
According to \eqref{1.1}, \eqref{1.1.1}, \eqref{1.2} and \eqref{1.2.1}, it can be easily shown that
\begin{align}
    \boldsymbol{h}_e\sim F_{h_e}\in\mathscr{D}(\bar{\boldsymbol{h}}_e+\boldsymbol{\xi}_{h_e},\boldsymbol{\Sigma}_{h_e}),\\
\boldsymbol{g}_e\sim F_{g_e}\in\mathscr{D}(\bar{\boldsymbol{g}}_e+\boldsymbol{\xi}_{g_e},\boldsymbol{\Sigma}_{g_e}).
\end{align}
To simplify the analysis, we define $\boldsymbol{\psi}_e=[\boldsymbol{h}_e^{\mathrm{T}},\boldsymbol{g}_e^{\mathrm{T}}]^{\mathrm{T}}\in \mathbb{C}^{(L+M)\times 1}$, and assume that $\boldsymbol{h}_e$ and $\boldsymbol{g}_e$ are mutually independent. Then, we can obtain
\begin{equation}
\boldsymbol{\psi}_e\sim F_e\in\mathscr{D}(\boldsymbol{\bar{\psi}}_e+\boldsymbol{\xi}_e,\boldsymbol{\Sigma}_e),
\end{equation}
where $\bar{\boldsymbol{\psi}}_e=[\bar{\boldsymbol{h}}_e^{\mathrm{T}},\bar{\boldsymbol{g}}_e^{\mathrm{T}}]^{\mathrm{T}}\in \mathbb{C}^{(L+M)\times 1}$, $\boldsymbol{\xi}_e=[\boldsymbol{\xi}^{\mathrm{T}}_{h_e},\boldsymbol{\xi}_{g_e}^{\mathrm{T}}]^{\mathrm{T}}$, and $\boldsymbol{\Sigma}_e=\mathrm{diag}(\boldsymbol{\Sigma}_{h_e},\boldsymbol{\Sigma}_{g_e})$. Meanwhile, let $\boldsymbol{V}=[\boldsymbol{I},\boldsymbol{G}^{\dagger}\boldsymbol{\Theta}^{\dagger}]$, the SINR of the $e$-th eavesdropper can be expressed as 
\begin{equation}
\gamma_{e,k}=\frac{\boldsymbol{\psi}^{\dagger}\boldsymbol{V}^{\dagger}\boldsymbol{w}_k\boldsymbol{w}_k^{\dagger}\boldsymbol{V}\boldsymbol{\psi}}{\sum\limits_{l=1,l\neq k}^K\boldsymbol{\psi}^{\dagger}\boldsymbol{V}^{\dagger}\boldsymbol{w}_l\boldsymbol{w}_l^{\dagger}\boldsymbol{V}\boldsymbol{\psi}+\sigma_e^2}, \quad\forall e\in\mathcal{E},\forall k\in\mathcal{K}.
\end{equation}

\section{Proposed Formulation and Decomposition}\label{s3.1}
\subsection{Proposed Formulation}
To ensure the security of satellite communication, we employ the ARIS to enhance legitimate transmission while minimizing information leakage to unintended parties. Consequently, the problem is formulated to maximize the total rate for all legitimate users, while limiting the rate of eavesdroppers by jointly optimizing the transmission beamforming and passive beamforming of the ARIS. Thus, the problem can be formulated as
\begin{subequations}\label{2.1.8}
\begin{align}
    \max_{\{\boldsymbol{w}_k\}_{k=1}^K,\boldsymbol{\theta}}\quad &\sum\limits_{k=1}^K\mathrm{log}(1+\gamma_k)\\
    \mathrm{s.t.}\quad&|\theta_m|=1,\quad\forall m\in \mathcal{M},\label{2.1.1}\\
    &\sum_{k=1}^K||\boldsymbol{w}_k||^2 \leq P_T,\label{2.1.7}\\
    &\begin{aligned}[t]\label{2.1.2}
      \inf_{F_e\in\mathscr{D}(\boldsymbol{\bar{\psi}}_e+\boldsymbol{\xi}_e,\boldsymbol{\Sigma}_e)}&\mathbb{P}_{\boldsymbol{\psi}_e\sim F_e}\{\mathrm{log}(1+\gamma_{e,k})\leq\Upsilon_k\}\\
      &\geq1-\epsilon_e,\quad\forall e\in\mathcal{E},\forall k\in\mathcal{K},
    \end{aligned}
\end{align}
\end{subequations}
where \eqref{2.1.1} represents the unit-modulus constraint on the ARIS phase shifters, $\Upsilon_{k}$ in \eqref{2.1.2} is the introduced threshold of wiretap rate for the $e$-th eavesdropper, and $\epsilon_e$ in  \eqref{2.1.2} denotes the probability that the SINR of all eavesdroppers remains below a prescribed threshold, valid for any probability distribution characterized by the specified mean and covariance. Notably, our model can be conveniently extended to address energy issues by reformulating the rate constraint into a data-volume constraint that reflects practical application requirements.

The formulated problem is rather complicated with threefold difficulties. First, the unit-modulus constraints imposed on the ARIS phase shifters are non-convex. Second, the objective function is non-concave, and the transmit beamforming and ARIS phase shifts are highly coupled, making the problem non-convex. Third, the security constraint \eqref{2.1.2} is a distributionally robust chance constraint, which is semi-infinite and generally intractable to handle directly.

\subsection{Uncertainty Treatment}

Given the non-convex nature of the distributionally robust chance constraint~\eqref{2.1.2}, it is necessary to transform it into a tractable convex constraint. Then, we first rewrite the chance constraint ~\eqref{2.1.2} as
\begin{equation}\label{eq:chance_rewrite}
\begin{aligned}
    \inf_{F_e\in\mathscr{D}(\boldsymbol{\bar{\psi}}_e+\boldsymbol{\xi}_e,\boldsymbol{\Sigma}_e)}&\mathbb{P}_{\boldsymbol{\psi}_e\sim F_e}\{L_{e,k}(\boldsymbol{\psi}_e,\boldsymbol{w}_k)\leq 0\}\geq1-\epsilon_e,
\\
&\forall e\in\mathcal{E},\forall k\in\mathcal{K},
\end{aligned}
\end{equation}
where $L_{e,k}(\boldsymbol{\psi}_e,\boldsymbol{w}_k) =\boldsymbol{\psi}_e^{\dagger}\boldsymbol{V}^{\dagger}(\boldsymbol{w}_k\boldsymbol{w}_k^{\dagger} - (2^{\Upsilon_k}-1)\sum\limits_{l=1,l\neq k}^K\allowbreak\boldsymbol{w}_l\boldsymbol{w}_l^{\dagger})\boldsymbol{V}\boldsymbol{\psi}_e - (2^{\Upsilon_k}-1)\sigma_e^2$ is quadratic in $\boldsymbol{\psi}_e$. Due to the difficulty in directly handling the worst-case probabilistic constraint, we introduce the Conditional Value-at-Risk (CVaR) as a convex approximation for the chance constraint. According to \cite[Theorem 2.2]{a3}, the constraint~\eqref{eq:chance_rewrite} can be transformed into
\begin{equation}\label{eq:cvar_equiv}
\begin{aligned}
    \sup_{F_e\in\mathscr{D}(\boldsymbol{\bar{\psi}}_e+\boldsymbol{\xi}_e,\boldsymbol{\Sigma}_e)}&\text{CVaR}_{\epsilon_e}(L_{e,k}(\boldsymbol{\psi}_e,\boldsymbol{w}_k)) \leq 0 ,\\
&\forall e\in\mathcal{E},\forall k\in\mathcal{K}.
\end{aligned}
\end{equation}
Based on the properties of CVaR, the aforementioned expression can be further expanded as
\begin{equation}\label{re1}
    \begin{aligned}
&\sup_{F_e\in\mathscr{D}}\text{CVaR}_{\epsilon_e}(L_{e,k}(\boldsymbol{\psi}_e,\boldsymbol{w}_k))\\
=&\sup_{F_e\in\mathscr{D}} \inf_{B_e\in\mathbb{R}} \left\{ B_e + \frac{1}{\epsilon_e} \mathbb{E}_{F_e} \left[ \left( L_{e,k}(\boldsymbol{\psi}_e,\boldsymbol{w}_k) - B_e \right)^+ \right] \right\}\\
=&B_e + \frac{1}{\epsilon_e} \sup_{F_e\in\mathscr{D}} \mathbb{E}_{F_e} \left[ \left( L_{e,k}(\boldsymbol{\psi}_e,\boldsymbol{w}_k) - B_e \right)^+ \right].
\end{aligned}
\end{equation}
In the last equality of~\eqref{re1}, the order of the maximization and minimization operations is interchanged by invoking the stochastic saddle point theorem proposed by Shapiro and Kleywegt~\cite{a4}. Furthermore, according to \cite[Lemma A.1]{a3}, the supremum in~\eqref{re1} is equivalent to
\begin{subequations}
\begin{align}
\min_{\boldsymbol{A}_{e,k} \in \mathbb{H}^{L+M+1}, \boldsymbol{A}_{e,k} \succeq 0} \quad & \mathrm{Tr}(\boldsymbol{\Omega}_e\boldsymbol{A}_{e,k}) \\
\mathrm{s.t.} \quad & \left[\boldsymbol{\psi}_e^{\dagger}, 1\right] \boldsymbol{A}_{e,k} \left[\begin{matrix}\boldsymbol{\psi}_e \\ 1\end{matrix}\right] \geq L_{e,k}(\boldsymbol{\psi}_e) - B_e,\nonumber\\
&\forall e\in\mathcal{E},\forall k\in\mathcal{K}.\label{re2} 
\end{align}
\end{subequations}
where $\boldsymbol{\Omega}_e=\left[\begin{matrix}
    \boldsymbol{\Sigma}_e+(\boldsymbol{\bar{\psi}_e}+\boldsymbol{\xi}_e)(\boldsymbol{\bar{\psi}_e}+\boldsymbol{\xi}_e)^{\dagger}&(\boldsymbol{\bar{\psi}_e}+\boldsymbol{\xi}_e)\\
    (\boldsymbol{\bar{\psi}_e}+\boldsymbol{\xi}_e)^{\dagger}&1
\end{matrix}\right]$. Since $L_{e,k}(\boldsymbol{\psi}_e,\boldsymbol{w}_k)$ is a quadratic in $\boldsymbol{\psi}_e$, the constraint~\eqref{re2} can be equivalently transformed into \eqref{2.1.4}, which is shown at the top of the next page. By introducing the auxiliary slack variables $\boldsymbol{A}_{e,k}$ and $B_e$, and substituting~\eqref{2.1.4} for \eqref{re2}, the original distributed robust chance constraint~\eqref{2.1.2} can be reformulated as
\begin{align}
     &  B_e+\frac{1}{\epsilon_e}\mathrm{Tr}(\boldsymbol{\Omega}_e\boldsymbol{A}_{e,k})\leq0,\quad\forall e\in\mathcal{E},\forall k\in\mathcal{K},\label{2.1.3}\\    
       &  \boldsymbol{A}_{e,k}\succeq0, \quad\forall e\in\mathcal{E}, \forall k\in\mathcal{K},\label{2.1.5}
\end{align}
and \eqref{2.1.4}, which is shown at the top of the next page.
\begin{figure*}[!t]
\begin{align}
&\boldsymbol{A}_{e,k}\succeq\left[\begin{matrix}\boldsymbol{V}^{\dagger}\left(\boldsymbol{w}_k\boldsymbol{w}_k^{\dagger}-(2^{\Upsilon_k}-1)\sum\limits_{l=1,l\neq k}^K\boldsymbol{w}_l\boldsymbol{w}_l^{\dagger}\right)\boldsymbol{V}&\boldsymbol{0}\\ \boldsymbol{0}&-(2^{\Upsilon_k}-1)\sigma_e^2-B_e\end{matrix}\right],\quad\forall e\in\mathcal{E},\forall k\in\mathcal{K}.\label{2.1.4}
\end{align}
\hrulefill
\end{figure*} 
Finally, the original problem \eqref{2.1.8} is reformulated as
\begin{subequations}\label{2.1.6}
\begin{align}
\max_{\boldsymbol{w}_k,\boldsymbol{\theta},\boldsymbol{A}_{e,k},B_e}\quad &\sum\limits_{k=1}^K\mathrm{log}(1+\gamma_k)\\
    \mathrm{s.t.}\quad&\eqref{2.1.1}, \eqref{2.1.7},\eqref{2.1.3}, \eqref{2.1.5}, \eqref{2.1.4}\nonumber.
\end{align}
\end{subequations}

The problem in \eqref{2.1.6} remains a non-convex optimization problem. Specifically, the transmit beamforming vector $\boldsymbol{w}_k$ and the ARIS phase shifts $\boldsymbol{\Theta}$ are coupled within both the objective function and constraint \eqref{2.1.4}. Additionally, the problem involves non-convex constraints, including the unit-modulus constraint \eqref{2.1.1} and the constraint \eqref{2.1.4}, which exhibit a non-convex quadratic difference structure. 

Due to these complexities, we utilize an AO framework to decompose it into two tractable sub-problems regarding transmission beamforming and passive beamforming. Specifically, the proposed framework alternately solves these two sub-problems through an iterative approach. We first optimize the transmission beamforming with fixed passive beamforming variables,  and subsequently optimize the passive beamforming with fixed transmit beamforming variables. The detailed reformulation and solution process for each sub-problem will be presented in the following subsections.

\section{Proposed Solution}\label{s3.2}
\subsection{Transmission Beamforming}
With the fixed passive beamforming, we formulate the subproblem for the transmission beamforming as
\begin{subequations}
\begin{align}
\max_{\boldsymbol{w}_k,\boldsymbol{A}_{e,k},B_e}\quad &\sum\limits_{k=1}^K\mathrm{log}(1+\gamma_k)\\
    \mathrm{s.t.}\quad& \eqref{2.1.7},\eqref{2.1.3}, \eqref{2.1.5},\eqref{2.1.4}.\nonumber
\end{align}
\end{subequations}
Given the non-convexity of the objective function with respect to the transmit beamforming vector $\boldsymbol{w}_k$, we simplify the notation by defining
\begin{equation}
\begin{aligned}
\boldsymbol{\Xi}_{k}=&\Bigg(\boldsymbol{h}^{\dagger}_{k}+\boldsymbol{g}^{\dagger}_{k}\boldsymbol{\Theta} \boldsymbol{G}\Bigg)\Bigg(\boldsymbol{h}_{k}^{\dagger}+\boldsymbol{g}^{\dagger}_{k}\boldsymbol{\Theta}\boldsymbol{G}\Bigg)^{\dagger},\quad\forall k\in\mathcal{K}.
\end{aligned}    
\end{equation}
Then, by defining $\boldsymbol{W}_k=\boldsymbol{w}_k{\boldsymbol{w}_k}^{\dagger}$, and the equivalent power constraint $\sum\limits_{k=1}^K\mathrm{Tr}(\boldsymbol{W}_k)\leq P_T$, the rates of legitimate users $R_{k}$ can be reformulated as
\begin{equation}
\begin{split}
R_{k}=&\log\left(\frac{\sum\limits_{l=1}^{K}\mathrm{Tr}(\boldsymbol{\Xi}_{k}\boldsymbol{W}_{l})+\sigma_{k}^{2}}{\sum\limits_{l=1,l\neq k}^K\mathrm{Tr}(\boldsymbol{\Xi}_{k}\boldsymbol{W}_l)+\sigma_{k}^{2}}\right)\\
=&\log\left(\sum_{l=1}^{K} \mathrm{Tr} (\boldsymbol{\Xi}_{k}\boldsymbol{W}_{l})+\sigma_{k}^{2}\right)\\
&-\log\left(\sum_{l=1,l\neq k}^K \mathrm{Tr} (\boldsymbol{\Xi}_{k}\boldsymbol{W}_{l})+\sigma_{k}^{2}\right),\\
&\forall k\in\mathcal{K}.
\end{split}
\end{equation}

For the objective function, based on~\cite[Lemma 1]{33}, we introduce the auxiliary variables $\xi_k$ to transform its non-convex form into
\begin{equation}
    \begin{aligned}R_k =&\log(\xi_{k})-\xi_{k}\left(\sum_{l=1,l\neq k}^{K} \mathrm{Tr} (\boldsymbol{\Xi}_{k}\boldsymbol{W}_{l})+\sigma_{k}^{2}\right)+1\\
&+\log\left(\sum_{l=1}^{K} \mathrm{Tr} (\boldsymbol{\Xi}_{k}\boldsymbol{W}_{l})+\sigma_{k}^{2}\right),\quad\forall k\in\mathcal{K}.
    \end{aligned}
\end{equation}
Meanwhile, the non-convex constraint \eqref{2.1.4} is recast as the following LMI \eqref{2.2.1} in terms of $\boldsymbol{W}_k$, which is shown at the top of the next page.
\begin{figure*}[t]
\begin{align}\label{2.2.1}
\boldsymbol{A}_{e,k}\succeq\left[\begin{matrix}\boldsymbol{V}^{\dagger}\left(\boldsymbol{W}_k-(2^{\Upsilon_k}-1)\sum\limits_{l=1,l\neq k}^K\boldsymbol{W}_l^{\dagger}\right)\boldsymbol{V}&\boldsymbol{0}\\\boldsymbol{0}&-(2^{\Upsilon_k}-1)\sigma_e^2-B_e\end{matrix}\right],\quad\forall e\in\mathcal{E},\forall k\in\mathcal{K}.
\end{align}
\hrulefill
\end{figure*}
Then, the transmission beamforming optimization problem becomes
\begin{subequations}\label{2.2.3}
\begin{align}   \max_{\boldsymbol{W}_k,\boldsymbol{A}_{e,k},B_e,\xi_k}\quad &\sum_{k=1}^{K}\Bigg(\log(\xi_{k})-\xi_{k}\Big(\sum_{l=1,l\neq k}^{K} \mathrm{Tr} (\boldsymbol{\Xi}_{k}\boldsymbol{W}_{l})\nonumber\\
&+\sigma_{k}^{2}\Big)+1+\log\Big(\sum_{l=1}^{K} \mathrm{Tr} (\boldsymbol{\Xi}_{k}\boldsymbol{W}_{l})+\sigma_{k}^{2}\Big)\Bigg)\\
    \mathrm{s.t.}\quad  &\xi_k\geq 0,\quad \forall k\in\mathcal{K},\\
    &\sum_{k=1}^K\mathrm{Tr}(\boldsymbol{W}_k) \leq P_T,\\
    &\boldsymbol{W}_k\succeq 0,\mathrm{rank}(\boldsymbol{W}_k)=1,\quad\forall k\in\mathcal{K},\label{2.2.2}\\
    &\eqref{2.1.3},\eqref{2.1.4},\eqref{2.2.1},\nonumber
\end{align}
\end{subequations}
where the rank-1 constraint specified in \eqref{2.2.2} is introduced by replacing $\{\boldsymbol{w}_k\}_{k=1}^K$ with $\{\boldsymbol{W}_k\}_{k=1}^K$. To solve this problem, an alternating optimization strategy is applied to the auxiliary variables and transmission beamforming. Specifically, the optimal $\xi_{k}$ is expressed as
\begin{equation}
    \xi_{k}=\frac{1}{\sum\limits_{l=1,l\neq k}^K\mathrm{Tr}(\boldsymbol{\Xi}_{k}\boldsymbol{W}_{l})+\sigma_{k}^{2}},\quad \forall k\in\mathcal{K}.
\end{equation}

Regarding the optimization variables $\{\boldsymbol{W}_{k}\}_{k=1}^K$, the application of semidefinite relaxation and the relaxation of the rank-1 constraint transform the formulation into a convex semidefinite program, which can be efficiently solved using standard toolboxes like CVX. The global solution for transmission beamforming is achieved by iteratively optimizing $\{\xi_{k}\}_{k=1}^K$ and $\{\boldsymbol{W}_{k}\}_{k=1}^K$ in \eqref{2.2.3}. Ultimately, the beamforming vector  $\boldsymbol{w}_{k}$ is extracted via eigenvalue decomposition of $\{\boldsymbol{W}_{k}\}_{k=1}^K$. Meanwhile, if the solution fails the rank-1 constraint, Gaussian Randomization is subsequently applied.

\subsection{Passive Beamforming of ARISs}
Given the fixed transmission beamforming, the passive beamforming subproblem is formulated as
\begin{subequations}\label{2.3.9}
\begin{align}
\max_{\boldsymbol{\theta},\boldsymbol{A}_{e,k},B_e}\quad &\sum\limits_{k=1}^K\mathrm{log}(1+\gamma_k)\label{2.3.12}\\
    \mathrm{s.t.}\quad&\eqref{2.1.1},\eqref{2.1.3},\eqref{2.1.5}, \eqref{2.1.4}.\nonumber
\end{align}
\end{subequations}
To facilitate the optimization of the RIS phase shifts, we first define the cascaded channel vector
\begin{align}\label{eq2}
&\boldsymbol{v}=[1,\boldsymbol{\vartheta}],\\
&\boldsymbol{\Lambda}_{k,l}=[\boldsymbol{h}_k^{\dagger}\boldsymbol{w}_l;\mathrm{diag}(\boldsymbol{g}_k^{\dagger})\boldsymbol{G}\boldsymbol{w}_l],\quad\forall l \in\mathcal{K}.
\end{align}   
Then, the rate of the $k$-th legitimate user can be reformulated into a compact form as
\begin{align}\label{2.3.13}
R_{k}=\log\left(1+\frac{\left|\boldsymbol{v}\boldsymbol{\Lambda}_{k,k}\right|^{2}}{\sum\limits_{l=1,l\neq k}^K\left|\boldsymbol{v}\boldsymbol{\Lambda}_{k,l}\right|^{2}+\sigma_{k}^{2}}\right),\quad\forall k\in\mathcal{K}.
\end{align}
It is readily observed that the objective function \eqref{2.3.12} represents the sum of achievable rates with fractional SINR expressions, resulting in a non-convex multiple-ratio fractional programming (FP) problem. Consequently, we leverage the quadratic transform technique~\cite{34} to decouple the numerator and denominator. By introducing the auxiliary vector $\boldsymbol{\alpha} \in \mathbb{C}^K$, where $\boldsymbol{\alpha} = [\alpha_1, \alpha_2, \dots, \alpha_K]$, the original rate expression in \eqref{2.3.13} is reformulated into
\begin{align}
 R_k =& \log\Big(1+2\mathrm{Re}\{\alpha_k^{\dagger}\boldsymbol{v}\boldsymbol{\Lambda}_{k,k}\}\nonumber\\
 &-\alpha_k^{\dagger}(\sum\limits_{l=1,l\neq k}^K|\boldsymbol{v}\boldsymbol\Lambda_{k,l}|^2+\sigma_k^2)\alpha_k\Big),\quad\forall k\in\mathcal{K}.
\end{align}

Meanwhile, to facilitate the handling of the non-convex constraint \eqref{2.1.4}, we define
\begin{align}
    &\boldsymbol{E}_k = \boldsymbol{G}\boldsymbol{w}_k\boldsymbol{w}_k^{\dagger}\boldsymbol{G}^{\dagger},\\
    &\boldsymbol{F}_k = (2^{\Upsilon_k}-1)\sum\limits_{l=1,l\neq k}^K\boldsymbol{G}\boldsymbol{w}_l\boldsymbol{w}_l^{\dagger}\boldsymbol{G}^{\dagger}.
\end{align}
Then, the constraint \eqref{2.1.4} is recast as \eqref{2.3.3}, which is shown at the top of the next page.
\begin{figure*}[t]
\begin{align}\label{2.3.3}
  &\boldsymbol{A}_{e,k}\succeq\left[\begin{matrix}\boldsymbol{w}_k\boldsymbol{w}_k^{\dagger}-(2^{\Upsilon_k}-1)\sum\limits_{l=1,l\neq k}^K\boldsymbol{w}_l\boldsymbol{w}_l^{\dagger}&\left(\boldsymbol{w}_k\boldsymbol{w}_k^{\dagger}-(2^{\Upsilon_k}-1)\sum\limits_{l=1,l\neq k}^K\boldsymbol{w}_l\boldsymbol{w}_l^{\dagger}\right)\boldsymbol{G}^{\dagger}\boldsymbol{\Theta}^{\dagger}&\boldsymbol{0}\\ \boldsymbol{\Theta}\boldsymbol{G}\left(\boldsymbol{w}_k\boldsymbol{w}_k^{\dagger}-(2^{\Upsilon_k}-1)\sum\limits_{l=1,l\neq k}^K\boldsymbol{w}_l\boldsymbol{w}_l^{\dagger}\right)&\boldsymbol{\Theta}\boldsymbol{E}_k\boldsymbol{\Theta}^{\dagger}-\boldsymbol{\Theta}\boldsymbol{F}_k\boldsymbol{\Theta}^{\dagger}&\boldsymbol{0}\\ \boldsymbol{0}&\boldsymbol{0}&-(2^{\Upsilon_k}-1)\sigma_e^2-B_e\end{matrix}\right],\nonumber\\
&\forall e\in\mathcal{E},\forall k\in\mathcal{K}.
\end{align}
\hrulefill
\end{figure*}
It is observed that the constraint in \eqref{2.3.3} is non-convex due to the indefinite quadratic terms coupled with the phase shift matrix $\boldsymbol{\Theta}$. Specifically, the term $\boldsymbol{\Theta}\boldsymbol{E}_k\boldsymbol{\Theta}^{\dagger} - \boldsymbol{\Theta}\boldsymbol{F}_k\boldsymbol{\Theta}^{\dagger}$ represents a difference-of-convex (DC) structure. To tackle the non-convexity in this constraint, we introduce a slack variable $\boldsymbol{Y}_k$ to upper-bound the quadratic term $\boldsymbol{\Theta}\boldsymbol{E}_k\boldsymbol{\Theta}^{\dagger}$ as
\begin{equation}
\boldsymbol{Y}_k\succeq\boldsymbol{\Theta}\boldsymbol{E}_k\boldsymbol{\Theta}^{\dagger}.
\end{equation}
Subsequently, by applying the Schur complement, we transform this inequality into an LMI given by
\begin{align}\label{2.3.11}
&\left[\begin{matrix}\boldsymbol{Y}_k&\boldsymbol{\Theta}\boldsymbol{G}\boldsymbol{w}_k\\\boldsymbol{w}_k^{\dagger}\boldsymbol{G}^{\dagger}\boldsymbol{\Theta}^{\dagger}&1
    \end{matrix}\right]\succeq0, \quad\forall k\in\mathcal{K}.
\end{align}
Meanwhile, the remaining term $-\boldsymbol{\Theta}\boldsymbol{F}_k\boldsymbol{\Theta}^{\dagger}$ can be linearized via its first-order Taylor expansion at $\boldsymbol{\Theta}^{(\ell)}$. Thus, the constraint \eqref{2.3.3} can be approximated as~\eqref{2.3.2}, which is shown at the top of the next page.
\begin{figure*}[t]
\begin{align}\label{2.3.2}
  &\boldsymbol{A}_{e,k}\succeq\left[\begin{matrix}\boldsymbol{
X
}_k&\boldsymbol{X}_k\boldsymbol{G}^{\dagger}\boldsymbol{\Theta}^{\dagger}&\boldsymbol{0}\\ \boldsymbol{\Theta}\boldsymbol{G}\boldsymbol{X}_k&\boldsymbol{Y}_k+\boldsymbol{\Theta}^{(\ell)}\boldsymbol{F}_{k}(\boldsymbol{\Theta}^{(\ell)})^{\dagger}-(\boldsymbol{\Theta}^{(\ell)})\boldsymbol{F}_{k}\boldsymbol{\Theta}^{\dagger}-\boldsymbol{\Theta}\boldsymbol{F}_{k}(\boldsymbol{\Theta}^{(\ell)})^{\dagger}&\boldsymbol{0}\\ \boldsymbol{0}&\boldsymbol{0}&-(2^{\Upsilon_k}-1)\sigma_e^2-B_e\end{matrix}\right],\quad\forall e\in\mathcal{E},\forall k\in\mathcal{K}.
\end{align}
\hrulefill
\end{figure*}

To address the non-convex unit-modulus constraint \eqref{2.1.1}, we first reformulate it as
\begin{align}
        |\theta_m|^2-1\leq 0,\quad\forall m\in\mathcal{M}, \label{2.3.4} \\
   1-|\theta_m|^2\leq 0,\quad\forall m\in\mathcal{M}. \label{2.3.5}
\end{align}
While \eqref{2.3.4} is convex, \eqref{2.3.5} represents a DC constraint, which motivates the adoption of the penalty Convex-Concave Procedure (CCP)~\cite{a5}. Specifically, we linearize the non-convex term $-|\theta_m|^2$ via a first-order Taylor expansion around the point $\theta_m^{(\ell)}$ from the $\ell$-th iteration, i.e.,
\begin{align}
    -|\theta_m|^2\leq|\theta_m^{(\ell)}|^2-2\mathrm{Re}\{(\theta_m^{(\ell)})^{\dagger}\theta_m\}\label{2.3.6},\quad\forall m\in\mathcal{M}.
\end{align}
Finally, problem \eqref{2.3.9} is recast as the following convex one:
\begin{subequations}\label{2.3.10}
\begin{align}
\max_{ \substack{ \boldsymbol{\theta}, \boldsymbol{A}_{e,k}, B_e, \\ \boldsymbol{\alpha}, \boldsymbol{a} \geq 0 } } \quad 
& \log\Big(1+2\mathrm{Re}\{\alpha_k^{\dagger}\boldsymbol{v}\boldsymbol{\Lambda}_{k,k}\}\\[-8pt] 
&-\alpha_k^{\dagger}(\sum\limits_{l=1,l\neq k}^K|\boldsymbol{v}\boldsymbol\Lambda_{k,l}|^2+\sigma_k^2)\alpha_k\Big)-\tau^{(\ell)}||\boldsymbol{a}||_1 \nonumber \\
\mathrm{s.t.} \quad 
& |\theta_m|^2\leq 1+a_m,\quad\forall m\in\mathcal{M}, \label{2.3.7}\\
& |\theta_m^{(\ell)}|^2 - 2\mathrm{Re}\left\{ (\theta_m^{(\ell)})^{\dagger} \theta_m \right\} \leq a_{M+m}-1, \nonumber \\
& \forall m\in\mathcal{M}, \label{2.3.8}\\
& \eqref{2.1.3},\eqref{2.1.5},\eqref{2.3.11},\eqref{2.3.2},\nonumber
\end{align}
\end{subequations}
where $\boldsymbol{a} = [a_1,\dots,a_{2M}]^{\mathrm{T}}$ denotes the vector of slack variables introduced to penalize violations of the unit-modulus constraints, and $\tau^{(\ell)}$ represents the penalty parameter at the $\ell$-th iteration. Specifically, the constraints in \eqref{2.3.11} and \eqref{2.3.2} represent the convex approximations of \eqref{2.1.4}, while \eqref{2.3.7} and \eqref{2.3.8} correspond to the convexified forms of \eqref{2.1.1}.  For this problem, we can similarly optimize the auxiliaries and passive beamforming alternately, where the optimized $\{\alpha_k\}_{k=1}^K$ can be obtained as
\begin{align}
    \alpha_k=\frac{|\boldsymbol{v}\boldsymbol{\Lambda}_{k,k}|^2}{\sum\limits_{l=1,l\neq k}^K|\boldsymbol{v}\boldsymbol{\Lambda}_{k,l}|^2},\quad\forall k\in\mathcal{K}.
\end{align}
Finally, the solution to the passive beamforming optimization problem can be obtained by iteratively optimizing $\{\alpha_k\}_{k=1}^K$ and $\boldsymbol{\theta}$ in \eqref{2.3.10}.
\subsection{Overall Algorithm Design}
\begin{algorithm}[!ht]\small
    \caption{Distributionally Robust Secure Beamforming Algorithm with ARIS}
    \renewcommand{\algorithmicrequire}{\textbf{Input:}}
    \renewcommand{\algorithmicensure}{\textbf{Output:}}
    \begin{algorithmic}[1]
       \Require Network topology, channel parameters
       \State Initialization: $t\leftarrow 0$, set $\boldsymbol{w}_k^{(t)},\boldsymbol{\theta}^{(t)};$
        \Repeat
            \State $t\leftarrow t+1;$
             \State Under the given $\boldsymbol{\theta}^{(t-1)}$;
                       \State  $\ell\leftarrow 0$;
                        \Repeat \Comment{SDR to update transmission beamforming}
                                    \State $\ell\leftarrow \ell+1;$
                                    \State Calculate $\xi_{k}^{(\ell)}$;
                                    \State Update $\boldsymbol{w}_k^{(\ell)}$ according to \eqref{2.2.3} with $\xi_{k}^{(\ell)}$;
                        \Until $\|\boldsymbol{w}_k^{(\ell)}-\boldsymbol{w}_k^{(\ell-1)}\|\leq\epsilon_T$;
                        \State$\boldsymbol{w}_k^{(t)}\leftarrow \boldsymbol{w}_k^{(\ell)}$;
    \State Under the given $\boldsymbol{w}_k^{(t)}$ ;
    \State $n\leftarrow 0$;
     \Repeat \Comment{FP and Penalty CCP to update passive beamforming}
            \State $n\leftarrow n+1$;
            \State Calculate $\alpha_{k}^{(n)}$;
             \State$\ell\leftarrow 0$;
            \Repeat
                        \State $\ell\leftarrow \ell+1$;
                        \State Update $\boldsymbol{\theta}^{(\ell)}$ according to \eqref{2.3.10} with $\alpha_k$ and obtain the dual variables $\{\kappa_m\}_{m=1}^{2M}$;
                        \State Update $\tau^{(\ell+1)}$: Set
                        \State
                         \begin{equation}
                            \tau^{(\ell+1)} = \begin{cases}
                                \tau^{(\ell)},\quad \mathrm{if}\space\tau^{(\ell)}>\Upsilon^{(\ell)}, \\
                                \mu\tau^{(\ell)},\quad \mathrm{if}\space \tau^{(\ell)}\leq\Upsilon^{(\ell)},
                            \end{cases}
                        \end{equation} 
                         \State  where 
                         \begin{equation}
                             \Upsilon^{(\ell)}\triangleq\min \left\{ \left\| \theta^{(\ell+1)} - \theta^{(\ell)} \right\|_2^{-1}, \max\limits_{m=1,\dots,2M} \left\{ \kappa_m \right\} \right\};
                         \end{equation}
             \Until $\|\boldsymbol{\theta}^{(\ell)}-\boldsymbol{\theta}^{(\ell-1)}\|\leq\epsilon_R$ and $||a||_1\leq\epsilon_P$;
             \State  $\boldsymbol{\theta}^{(n)}\leftarrow \boldsymbol{\theta}^{(\ell)}$;
             \Until $\|\boldsymbol{\theta}^{(n)}-\boldsymbol{\theta}^{(n-1)}\|\leq\epsilon_R$;
             \State  $\boldsymbol{\theta}^{(t)}\leftarrow \boldsymbol{\theta}^{(n)}$;
\Until{Convergence with the threshold $\epsilon_A$}
        \Ensure Transmission beamforming, passive beamforming of ARIS.
    \end{algorithmic}
\end{algorithm}
Based on the above discussion, we first address the distributionally robust chance constraints by introducing auxiliary variables and reformulating them into tractable forms. Then, we decompose the original optimization problem into transmission beamforming and passive beamforming subproblems. The solution to the reformulation of problem \eqref{2.1.6} can be obtained by solving these subproblems iteratively within an AO framework. Therefore, the proposed distributionally robust secure beamforming algorithm is summarized in Algorithm 1, where $t=0,1,2,\dots,$ denotes the outer iterations and the $\epsilon_A$ determines the overall convergence. The transmission beamforming requires an inner loop based on SDR, where the constant $\epsilon_T$ indicates the convergence. The passive beamforming requires inner loops utilizing FP and the CCP, where the $\epsilon_R$ indicates the convergence of the phase shifts and $\epsilon_P$ guarantees the satisfaction of the unit-modulus constraints. Furthermore, to prevent the algorithm from becoming trapped in an infeasible region, a small penalty parameter $\tau$ is typically initialized in the early stages of iteration and is dynamically increased according to a predetermined schedule. This strategy enables the CCP algorithm to prioritize the gradual improvement of the objective value during the initial phase, while progressively shifting the focus toward strictly satisfying the penalty constraints in the later stages. Since the transmit power is limited and the objective function is bounded, the proposed algorithm is guaranteed to converge.

In Algorithm 1, lines 6-10 correspond to the transmission beamforming subproblem. Meanwhile, lines 14-28 correspond to the passive beamforming subproblem with the fixed transmission beamforming. Notably, the proposed AO framework is specifically designed to address the unique mathematical complexities of the DRO formulation. The FP and penalty CCP are essential to resolve the DRO-induced difference-of-convex structure, while SDR explicitly handles the non-convex rank-1 constraints in the transmit beamforming subproblem.

Moreover, we briefly analyze the computational complexity of the proposed algorithm. For the transmission beamforming, solving the semidefinite programming has a complexity of $\mathcal{O}((K L^2 + E K (L+M)^2)^{3.5})$ and we assume there are $I_1$ times of iterations, then the incurred complexity is $\mathcal{O}(I_{1}(K L^2 + E K (L+M)^2)^{3.5})$. For the passive beamforming subproblem, the optimization procedure employs a double-loop structure comprising an outer FP loop and an inner penalty CCP loop. Solving the convexified problem incurs a computational complexity of $\mathcal{O}((M + E K (L+M)^2)^{3.5})$. Assuming the FP and CCP loops require $I_{2}$ and $I_{3}$ iterations, respectively, the complexity of the passive beamforming subproblem is $\mathcal{O}(I_{2} I_{3} (M + E K (L+M)^2)^{3.5})$. Finally, assuming the number of outer iterations in the AO framework is $I_0$, then the overall complexity is given by $\mathcal{O}(I_{0}[I_{1}(K L^2 + E K (L+M)^2)^{3.5} + I_{2} I_{3} (M + E K (L+M)^2)^{3.5}])$. Furthermore, regarding system scalability, although the computational complexity scales polynomially and security constraints become more stringent as the number of users and eavesdroppers increases, the proposed framework remains highly scalable in typical multibeam satellite scenarios.

\section{Simulation Results}\label{s4}
This section presents numerical results to demonstrate the performance of the proposed robust ARIS-assisted satellite system. We consider a downlink scenario where a Low Earth Orbit (LEO) satellite at an altitude of 220 km. The communication takes a carrier frequency of 6 GHz, and the Doppler effect caused by satellite movements is assumed to be compensated. The processing bandwidth is set to $B= 200$ MHz, and the noise temperature is $T_{i_k}= 290$ K, $\forall i\in\mathcal{I}$. There are $K=3$ legitimate users in the presence of $E=2$ eavesdroppers. The satellite is equipped with $L=5$ antennas and operates with a total transmission power budget of $P_T= 100$ W. Following the European Space Agency (ESA) model~\cite{35}, the maximum antenna gain is set to 50 dBi. The ARIS is deployed at an altitude of 100 m. To compensate for significant path loss, each ARIS comprises 15 subsurfaces, with each sub-surface containing 15 reflecting elements spaced by half a wavelength. The reference path loss at 1 m is 20 dB, and the air-to-ground path loss exponent is 2.3. Finally, security constraints are enforced by setting the eavesdropper rate threshold to 0.3 bps/Hz and the outage probability threshold to $\epsilon = 0.1$. The parameters of Algorithm 1 are chosen as $\tau^{(0)} = 0.001$, $\mu = 2$ and $\epsilon_A = \epsilon_T = \epsilon_R = \epsilon_P = 1 \times 10^{-3}$. The hyperparameter $\tau^{(0)}$ is initialized to a sufficiently small value to prevent the algorithm from becoming trapped in an infeasible region, thereby enabling the CCP algorithm to prioritize the gradual improvement of the objective value during the initial phase. The scaling factor $\mu$ is adopted to balance the convergence speed and the strict satisfaction of the penalty constraints in the later stages of the iterations.

For the eavesdroppers' CSI error model, we set the mean vectors $\xi_{h_e} = -0.05 \times \frac{\lambda}{4\pi d_{s,e}} \times (1_L + j1_L)$, $\xi_{g_e} = -0.05 \times C_L(d_{a,e}) \times (1_M + j1_M)$ and the covariances $\Sigma_{h_e} = 0.01 \times \frac{\lambda}{4\pi d_{s,e}} \times I_L$, $\Sigma_{g_e} = 0.01 \times C_L(d_{a,e}) \times I_M$, where $d_{s,e}$ and $d_{a,e}$ denotes the distance from the satellite to the e-th eavesdropper and from the ARIS to the e-th eavesdropper, respectively. Specifically, the covariance matrices are scaled according to the large-scale path loss of their respective links to ensure that the error variance remains physically consistent with the actual channel attenuation.

In the following simulations, we adopt two benchmark schemes: ``DRO without RIS'' and ``Perfect CSI". For the benchmark of "DRO without RIS", there is no ARIS in the system and communication relies on the direct link. The transmission beamforming at the satellite is optimized using the proposed DRO algorithm considering the CSI uncertainty of the direct links only.  For the "Perfect CSI" benchmark, the beamforming is optimized by assuming that the estimated CSI($\boldsymbol{\bar{\psi}}_e$) is perfect, i.e., $\boldsymbol{\xi}_e = \mathbf{0}$. It disregards the CSI uncertainty constraint \eqref{2.1.2} and solves the original problem directly.
\begin{figure*}[t]
    \centering
    \includegraphics[width=0.6\linewidth]{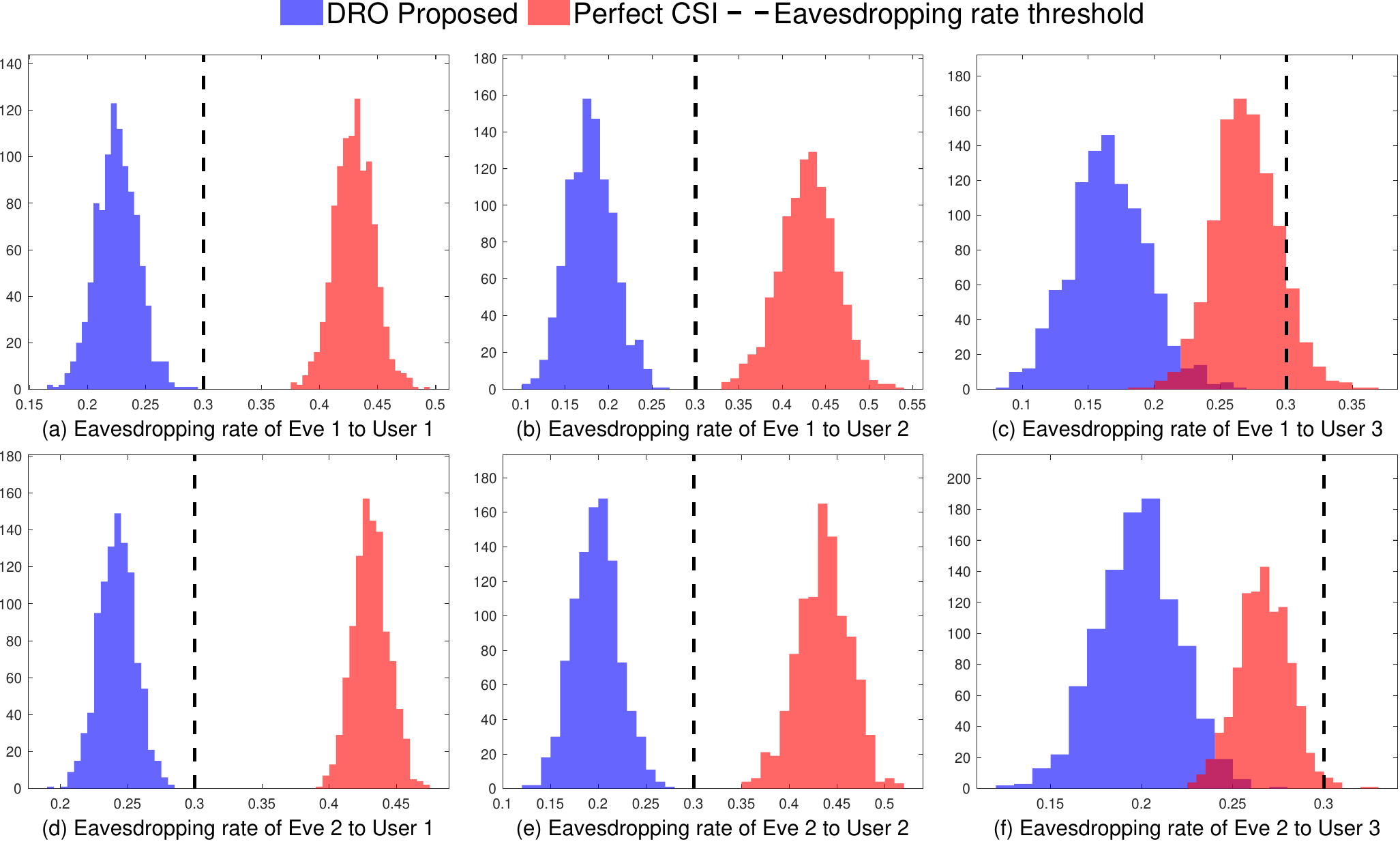}
    \caption{Histograms of Eves' rates for the different legitimate users.}
    \label{fig1}
\end{figure*}
Fig. \ref{fig1} presents the histograms of eavesdropping rates across different Eve-User links to validate the robustness of the proposed scheme against CSI uncertainties. It is evident that the eavesdropping rates achieved by the proposed DRO scheme are strictly below the preset threshold across all subfigures. This confirms that the proposed algorithm effectively mitigates information leakage by compensating for channel estimation errors. In contrast, the benchmark scheme based on perfect CSI significantly violates the security constraint, with the majority of its rate distribution exceeding the threshold. This performance degradation underscores the vulnerability of non-robust designs to imperfect CSI and highlights the necessity of the proposed DRO framework in guaranteeing physical layer security under CSI uncertainty.
\begin{figure*}[t]
    \centering
    \includegraphics[width=0.7\linewidth]{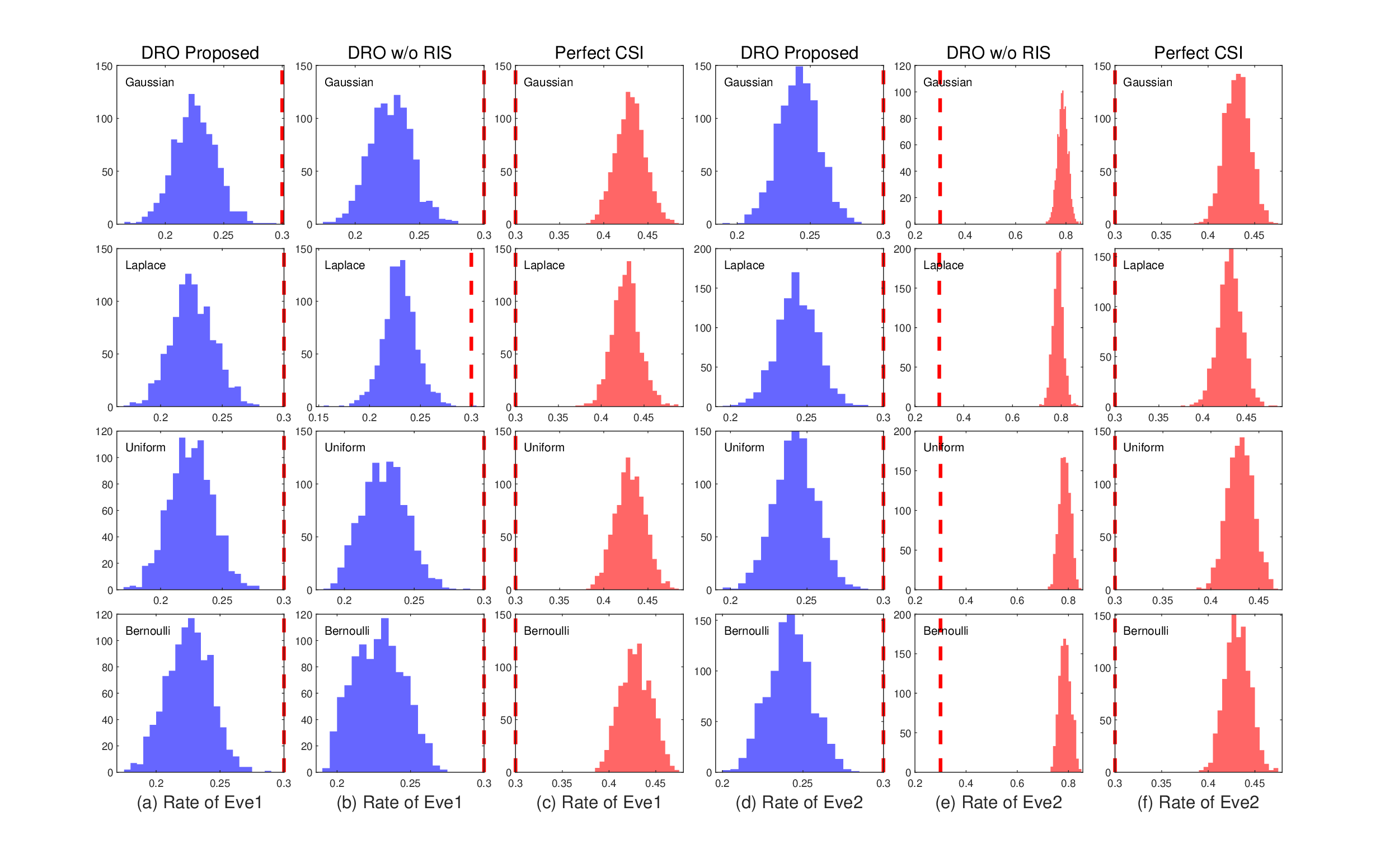}
    \caption{Histograms of Eves' rates under different distributions.}
    \label{fig2}
\end{figure*}

To examine the robustness of the proposed algorithm, Fig. \ref{fig2} presents the rate histograms of the eavesdroppers with respect to the same legitimate user under different CSI error distributions. Given that the distributions satisfying $\mathscr{D}(\boldsymbol{\bar{\psi}}_e+\boldsymbol{\xi}_e,\boldsymbol{\Sigma}_e)$ take various forms, we investigate four representative distributions: Gaussian, Binary, Uniform, and Laplacian. We design the corresponding beamforming schemes based on the proposed DRO algorithm, the DRO algorithm without ARIS, and the algorithm based on perfect CSI, respectively. In the simulations, $10^3$ error samples $\Delta \mathbf{h}_e$ ($\Delta \mathbf{g}_e$) are randomly generated for each distribution to obtain the actual channels $\mathbf{h}_e$ ($\mathbf{g}_e$). Subsequently, the rates of Eve 1 and Eve 2 are calculated under the different channel error distributions and beamforming algorithms. In the figure, the red dashed line denotes the maximum allowable rate threshold $\gamma_e$, while the region to the right of the dashed line corresponds to outage events. Specifically, Figs. \ref{fig2}(a) and (d) illustrate the performance of the proposed DRO scheme, Figs. \ref{fig2}(b) and (e) correspond to the DRO scheme without ARIS and Figs. \ref{fig2}(c) and (f) display the results designed based on perfect CSI. It can be observed from Fig. \ref{fig2} that the proposed DRO algorithm ensures that the rates of all eavesdroppers remain below $\gamma_e$ under all four error distributions, implying that no outage occurs. In contrast, although the scheme without ARIS satisfies the constraint for Eve 1, it results in outages for Eve 2, which demonstrates the necessity of introducing ARIS. Furthermore, the algorithm designed based on perfect CSI experiences outages under all distributions, indicating that non-robust designs violate constraint \eqref{2.1.2} to varying degrees.
\begin{figure}[t]
    \centering
    \includegraphics[width=0.8\linewidth]{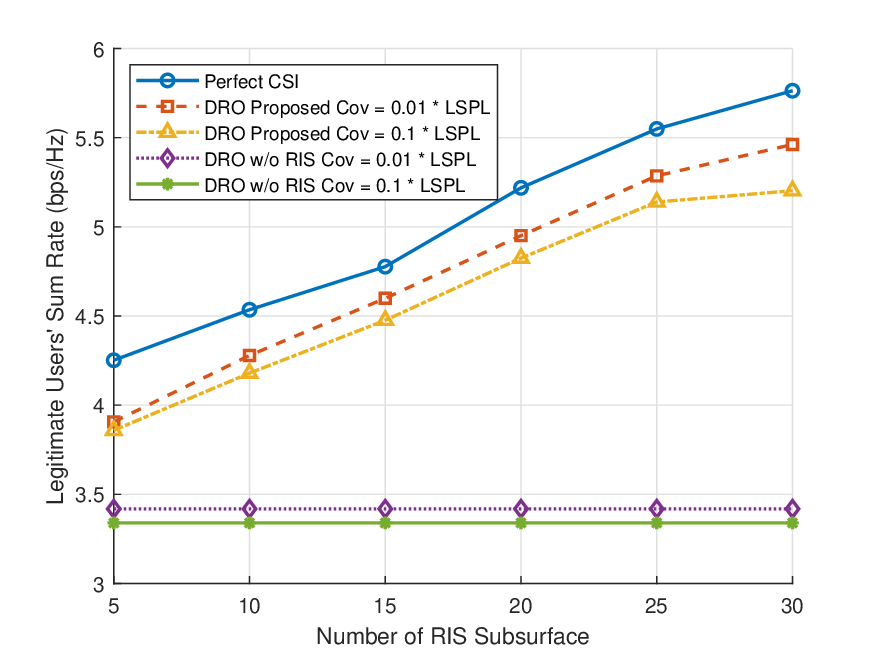}
    \caption{Legitimate users' sum rate versus the number of ARIS subsurfaces.}
    \label{fig3}
\end{figure}

Fig. \ref{fig3} illustrates that the legitimate users' sum rate increases monotonically with the number of ARIS subsurfaces. This substantial improvement is primarily attributed to the increased degrees of freedom. Specifically, a larger number of subsurfaces enables more flexible phase control to enhance the signal at the legitimate receiver while simultaneously suppressing the signal at the eavesdropper. In contrast, the sum rate of the benchmarks without RIS is significantly inferior to that of the proposed ARIS scheme. This comparison confirms that the auxiliary links established via ARIS are indispensable for expanding the feasible region of secure beamforming.
\begin{figure}[t]
    \centering
    \includegraphics[width=0.8\linewidth]{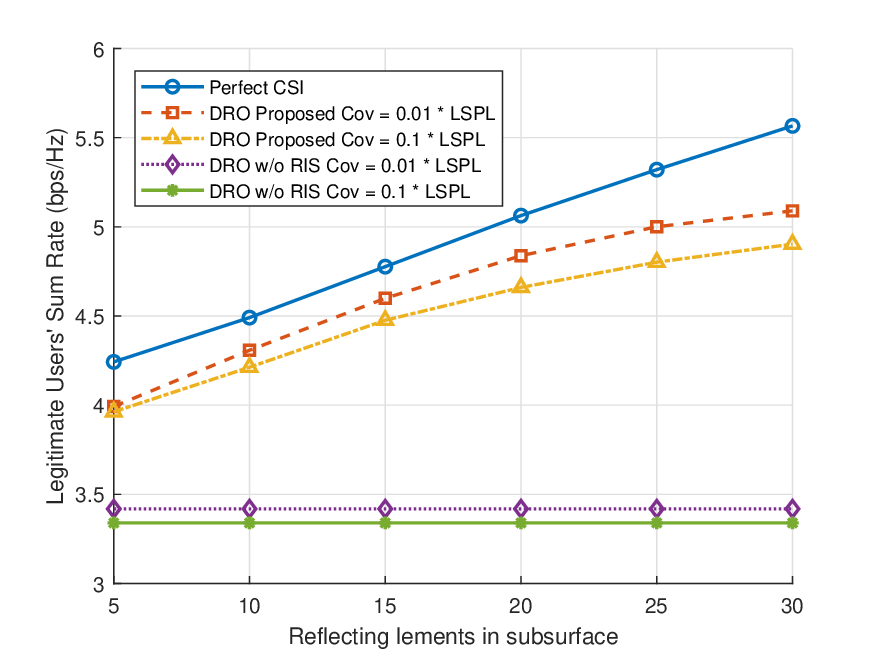}
    \caption{Legitimate users' sum rate versus the number of reflecting elements in each subsurface.}
    \label{fig4}
\end{figure}

Fig. \ref{fig4} presents the legitimate users' sum rate versus the number of reflecting elements within each subsurface. The performance of the proposed DRO scheme improves as the number of reflecting elements within the subsurfaces increases. This improvement is attributed to the enhanced passive beamforming gain provided by the additional elements. Conversely, the scheme without RIS lacks the aperture gain provided by the subsurfaces, resulting in a lower legitimate user rate. This performance gap indicates the necessity of deploying subsurfaces to compensate for long-distance satellite signal attenuation. Consequently, the results confirm that increasing the number of reconfigurable elements within each subsurface effectively improves the achievable rate of legitimate users in the presence of eavesdroppers.
\begin{figure}[t]
    \centering
    \includegraphics[ width=0.8\linewidth]{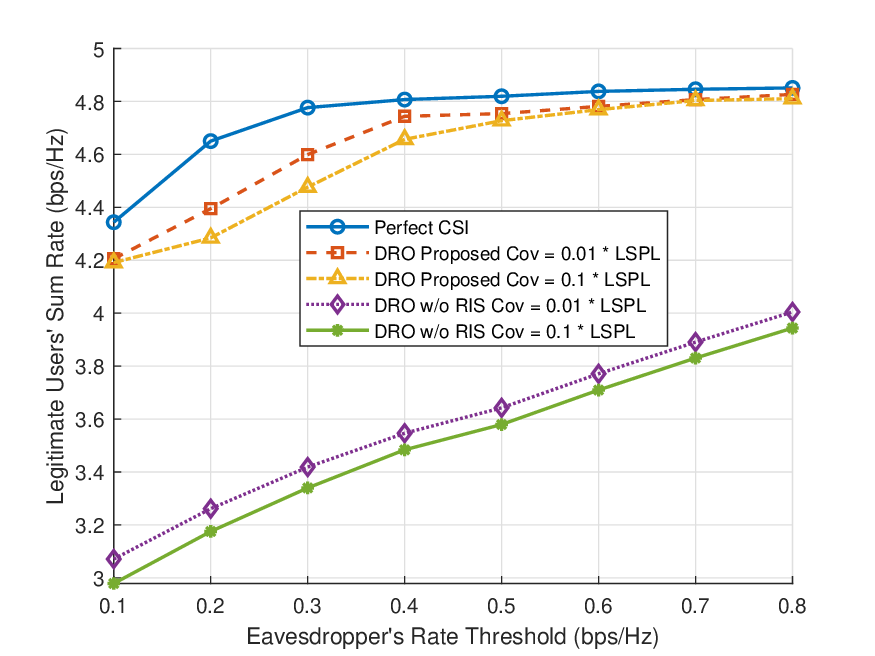}
    \caption{Legitimate users' sum rate versus the eavesdropper's rate threshold.}
    \label{fig5}
\end{figure}

Fig. \ref{fig5} depicts the legitimate users' sum rate versus the eavesdropper's rate threshold. It is evident that the sum rate increases monotonically as the eavesdropping rate threshold is relaxed. This is because the optimization algorithm is less constrained in suppressing signal leakage, allowing more power to be allocated toward maximizing legitimate users' rates. Notably, while the benchmarks without RIS continue to rise, the proposed RIS-assisted schemes and the Perfect CSI case reach a plateau when the threshold exceeds 0.5 bps/Hz. This saturation implies that the maximum achievable rate for the eavesdropper is approximately 0.5 bps/Hz under the current geometric configuration. Furthermore, the performance of the proposed robust design gradually approaches the Perfect CSI upper bound in this region, indicating that the impact of channel uncertainty diminishes when secrecy constraints are relaxed.

\begin{figure}[t]
    \centering
    \includegraphics[ width=0.8\linewidth]{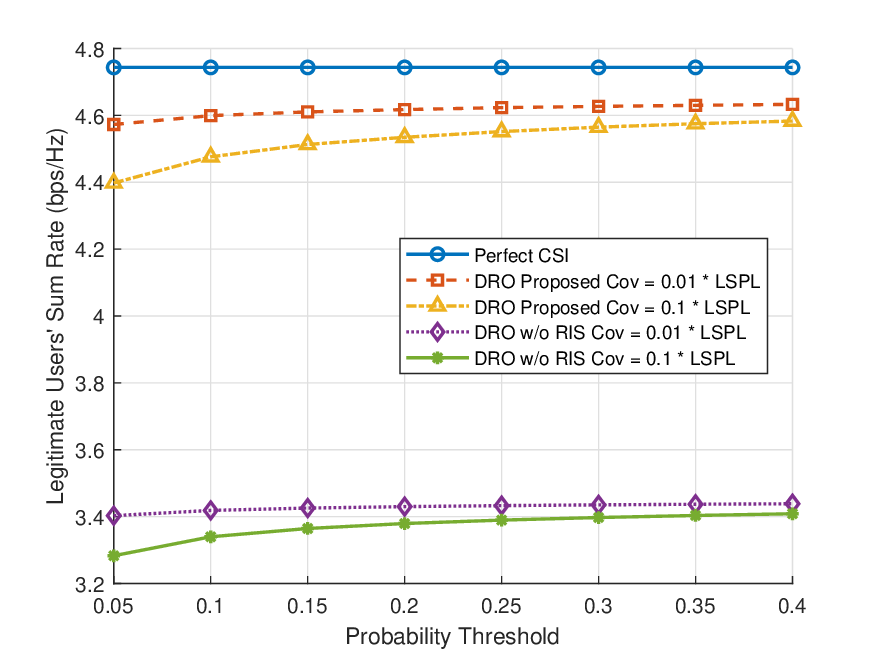}
    \caption{Legitimate users' sum rate versus the outage probability threshold.}
    \label{fig6}
\end{figure}

Fig. \ref{fig6} depicts the legitimate users' achievable sum rate versus the outage probability threshold. The sum rate for all schemes increases monotonically with $\epsilon$. This trend arises because a relaxed secrecy outage constraint enables the satellite to prioritize power allocation for information transmission over strict leakage suppression. Moreover, the proposed DRO scheme exhibits better performance than the "DRO w/o RIS" benchmark, which confirms the effectiveness of the ARIS in reconfiguring the propagation environment to enhance secrecy performance. Furthermore, the scenario with lower error covariance (i.e., 0.01) yields a higher rate compared to the larger covariance case (i.e., 0.1), indicating that superior CSI precision facilitates more accurate beamforming.
\begin{figure}[t]
    \centering
    \includegraphics[width=0.8\linewidth]{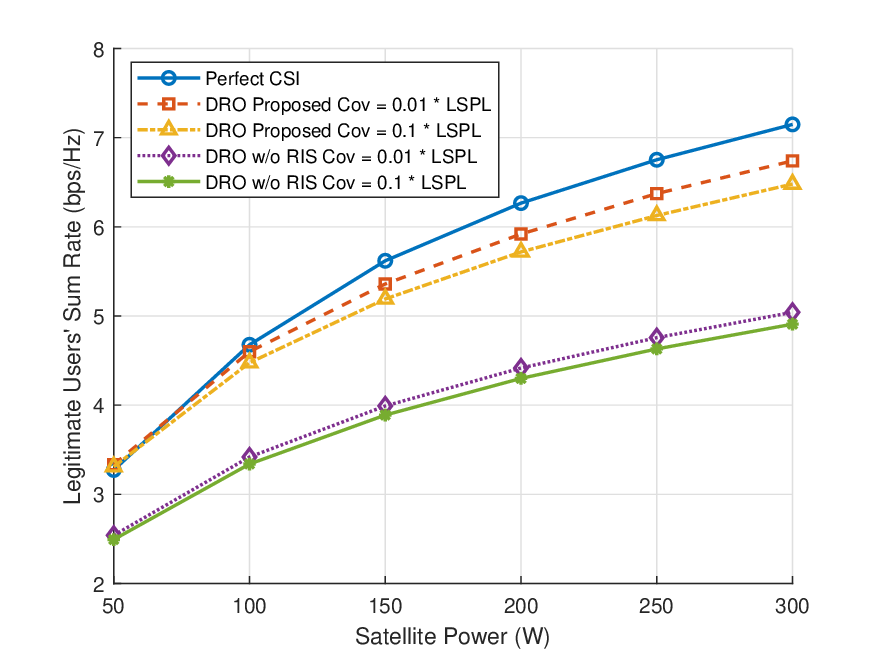}
    \caption{Legitimate users' sum rate versus the satellite transmit power.}
    \label{fig7}
\end{figure}

Fig. \ref{fig7} presents the legitimate users' sum rate versus the satellite transmit power. As expected, the sum rate for all considered schemes increases with the transmit power. However, the proposed DRO scheme with ARIS exhibits more rapid growth and superior performance compared to the benchmark without ARIS. This indicates that the ARIS-assisted system is more power-efficient, capable of translating the increased power budget into useful signal strength for legitimate users more effectively by suppressing the interference at the eavesdroppers through robust passive beamforming. Notably, as the power increases, the robust design ensures that the improved transmit power does not translate into leakage to eavesdroppers, maintaining the outage probability constraint while maximizing the legitimate rate.
\begin{figure}[t]
    \centering
    \includegraphics[width=0.8\linewidth]{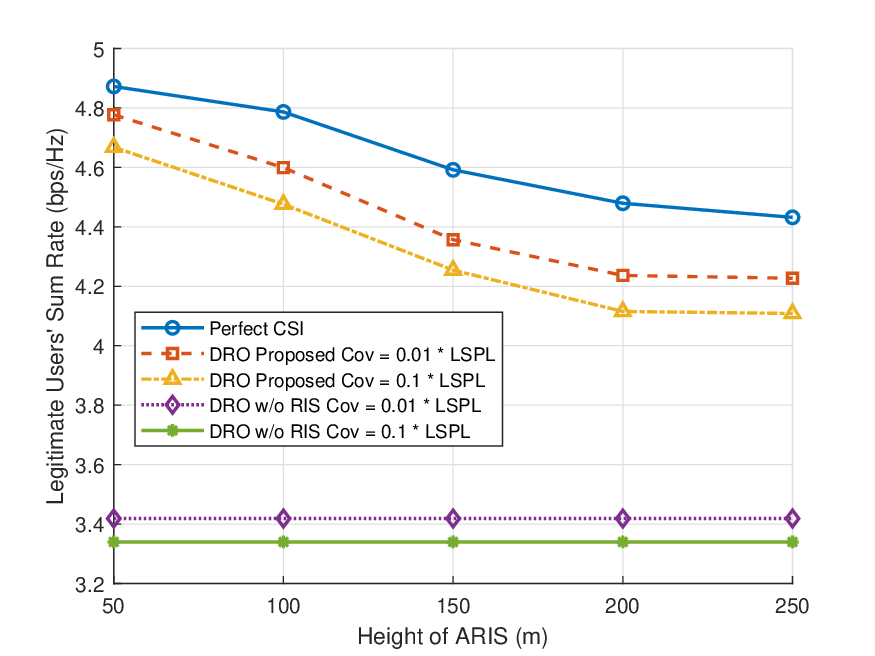}
    \caption{Legitimate users' sum rate versus the altitude of ARIS.}
    \label{fig8}
\end{figure}

Fig. \ref{fig8} illustrates the impact of the ARIS hovering altitude on the legitimate users' sum rate. The sum rate decreases monotonically as the ARIS altitude rises from 50 m to 250 m. It is observed that the performance of the proposed DRO scheme improves as $M$ increases, which is attributed to the enhanced passive beamforming gain provided by the reflecting units. Although a higher altitude enhances the LoS probability, the path loss induced by the extended propagation distance outweighs the benefits of improved channel conditions, resulting in a lower achievable rate. Nevertheless, the proposed robust design maintains a performance advantage over the benchmark schemes across varying altitudes, demonstrating its effectiveness in enhancing the legitimate users' rate.

\section{Conclusion}\label{s5}
In this paper, we have investigated a robust secure transmission strategy for ARIS-assisted multi-beam satellite communication systems. To address the inherent difficulty in obtaining accurate eavesdroppers' CSI, we adopted a distributionally robust optimization approach that models CSI errors using moment-based uncertainty sets rather than assuming a fixed probability distribution. Then, we formulated a joint beamforming optimization problem to maximize the legitimate users' sum rate while strictly limiting the leakage rate under a worst-case outage probability constraint. To solve the resulting non-convex optimization problem, we transform the intractable chance constraints into LMIs and propose an AO algorithm. Simulation results verified that the proposed scheme provides substantial security gains over systems without ARIS and maintains robustness across varying channel error distributions. These findings offer an effective solution for safeguarding future satellite networks. Furthermore, although this paper assumes a stationary ARIS, extending the model to a mobile aerial platform would enable dynamic trajectory optimization, thereby enhancing the quality of legitimate links and effectively suppressing eavesdropping. For such highly dynamic mobile aerial platforms in future low-altitude economy networks, extending the proposed framework to incorporate multi-agent learning-based temporal spectrum cartography~\cite{b1} and security-aware joint sensing, communication, and computing optimization~\cite{b2} represent promising directions for future research.

\bibliographystyle{IEEEtran}
\bibliography{dro}

\end{document}